\documentclass[showpacs,prd,showkeys,aps,twocolumn]{revtex4-1}

\usepackage[latin9]{inputenc}
\usepackage{textcomp}
\usepackage{amsmath}
\usepackage{amssymb}
\usepackage{xcolor}

\begin{document}

\title{Eruptive Massive Vector Particles of 5-Dimensional Kerr-Gödel Spacetime}
\author{A. \"{O}vg\"{u}n}
\email{ali.ovgun@pucv.cl}
\affiliation{Instituto de F\'{\i}sica, Pontificia Universidad Cat\'olica de
Valpara\'{\i}so, Casilla 4950, Valpara\'{\i}so, Chile}
\affiliation{Physics Department, Arts and Sciences Faculty, Eastern Mediterranean University, Famagusta, North Cyprus via Mersin 10, Turkey}

\author{I. Sakalli}
\email{izzet.sakalli@emu.edu.tr}
\affiliation{Physics Department, Arts and Sciences Faculty, Eastern Mediterranean University, Famagusta, North Cyprus via Mersin 10, Turkey}

\date{\today }

\begin{abstract}
In this paper, we investigate Hawking radiation of massive spin-1 particles from 5-dimensional Kerr-Gödel spacetime. By applying the WKB approximation and
the Hamilton-Jacobi ansatz to the relativistic Proca equation, we obtain the
quantum tunneling rate of the massive vector particles. Using the obtained
tunneling rate, we show how one impeccably computes the Hawking temperature of the
5-dimensional Kerr-Gödel spacetime.
\end{abstract}

\maketitle

\section{Introduction}

Today, there are numerous methods that confirm the outstanding discovery of Hawking \cite{hawking1,hawking2,hawking3}, which theoretically showed that
a black hole can sparkle as such in a blackbody radiation. In the last decade, semi-classical methods of modeling Hawking radiation as a tunneling
process have garnered lots of interest \cite{kraus1,kraus2,perkih1,perkih2,perkih3,ang,sri1,sri2,Akhmedova0,Akhmedova1,Akhmedova2,Akhmedova3,Akhmedova4,mann0,mann1,borun,yale,singleton1,singleton2,rahman,md2,md3,md4,md5,md6,vanzo,mann2,mann3,mann4,qgv1,qgv2,nasser1,khalil2,a1,a2,a22,a222,a2222,a33,k6,xiang2,h1,h2,h3,h4}. Those
semi-classical methods model the Hawking radiation as if a tunneling process, which
uses the WKB approximation to compute the imaginary part of the
classically unpermitted action ($ImS$) of the trajectory that crosses the
event horizon. By this way, the tunneling probability is computed via the expression of $
\Gamma \propto exp(-2ImS).$

The studies about the Hawking radiation of massive vector particles have been
gained attention as from the publications of Kruglov \cite{kruglov1,kruglov2}. His main approach is to apply the Hamilton-Jacobi method to the Proca equation \cite{proca1}. As is
well known, the vector particles (e.g. $Z,W$\textpm)
play important role in the high energy physics and in the standard model \cite{wz1,wz2}. Since the Hawking quanta
should be highly energetic tiny particles, so it is plausible to consider the
Hawking radiation of the massive vector particles from various geometries of black
holes. A reader would appreciate that the most subtle black holes are the
rotating ones defined in the higher dimensions. From this viewpoint, we
aim to extend the Kruglov's studies (and his followers' works \cite{
 a3,a4,a5,s1,k1,k2,a6,k3,a7,k7,k4,k5}) to the radiation spectrum of massive vector particles
spraying from a 5-dimensional Kerr-Gödel black hole (5DKGBH). Thus, we aim to
calculate Hawking temperature \cite{wald} of the 5DKGBH.

One of the splendid exact solutions of Einstein equations with a
cosmological constant and homogeneous pressure-free matter is the Gödel
universe \cite{godel}. Gödel spacetime possesses a peculiar feature:
there always exist closed timelike or null curves. On
the other hand, the generic
version of the Gödel spacetime, which is the exact solution of minimal
supergravity (preserving some number of supersymmetries) in 5 dimensions was produced in \cite{gauntlett,herdeiro}. The consistency of this solution to the string theory was further studied in 
\cite{boyda,harmark} in which the supersymmetric Gödel
universes of \cite{gauntlett,herdeiro} are shown to be related with the
pp-wave solutions having $T$-duality \cite{blau,bekenstein}.

The organization of the paper is as follows. In Sec. II, we introduce the 5DKGBH spacetime and demonstrate its thermodynamic features.
We then study the quantum tunneling of massive spin-1 particles from the 5DKGBH
by employing the Proca equation. In sequel, we compute the quantum tunneling
rate and obtain the corresponding temperature of the black hole surface. In
Sec. III, we summarize our discussions.

\section{Kerr-Gödel Black Hole in 5-Dimensions}

In Boyer-Lindquist coordinates, the 5DKGBH spacetime is given by \cite{kerrgodel} 
\begin{equation}
ds^{2}=-f(r)(dt+\frac{a(r)}{f(r)}\sigma _{3})^{2}+\frac{dr^{2}}{V(r)}+\frac{%
r^{2}}{4}(\sigma _{1}^{2}+\sigma _{2}^{2})+\frac{r^{2}V(r)}{4f(r)}\sigma
_{3}^{2},  \label{1}
\end{equation}%
where 
\begin{eqnarray}
f(r) &=&1-\frac{2m}{r^{2}},  \label{2} \\
a(r) &=&jr^{2}+\frac{ml}{r^{2}},  \label{3} \\
V(r) &=&1-\frac{2m}{r^{2}}+\frac{16j^{2}m^{2}}{r^{2}}+\frac{8jml}{r^{2}}+%
\frac{2ml^{2}}{r^{4}}.  \label{4}
\end{eqnarray}%
In the above equations, each $\sigma $ corresponds to the right-invariant
one-forms of SU(2) with the Euler angles ($\theta ,\phi ,\psi $). Namely, we
have 
\begin{eqnarray}
\sigma _{1} &=&\sin \phi d\theta -\cos \phi \sin \theta d\psi ,  \label{5} \\
\sigma _{2} &=&\cos \phi d\theta +\sin \phi \sin \theta d\psi ,  \label{6} \\
\sigma _{3} &=&d\phi +\cos \theta d\psi .  \label{7}
\end{eqnarray}%
It is possible to rewrite the metric \eqref{1} in different forms \cite{mann4}. 

(i) Expanded form: 
\begin{equation}
ds^{2}=-f(r)dt^{2}-2a(r)dt\sigma _{3}+g(r)\sigma _{3}^{2}+\frac{dr^{2}}{V(r)}%
+\frac{r^{2}}{4}(\sigma _{1}^{2}+\sigma _{2}^{2}),  \label{8}
\end{equation}%
where 
\begin{equation}
g(r)=\frac{r^{2}V(r)-4a^{2}(r)}{4f(r)}=-j^{2}r^{4}+\frac{1-8mj^{2}}{4}r^{2}+%
\frac{ml^{2}}{2r^{2}}.  \label{9}
\end{equation}%
(ii) Lapse-shift form:  
\begin{equation}
ds^{2}=-N^{2}dt^{2}+g(r)(\sigma _{3}-\frac{a(r)}{g(r)}dt)^{2}+\frac{dr^{2}}{%
V(r)}+\frac{r^{2}}{4}(\sigma _{1}^{2}+\sigma _{2}^{2}),  \label{10}
\end{equation}%
where 
\begin{equation}
N^{2}=f(r)+\frac{a^{2}(r)}{g(r)}=\frac{r^{2}V(r)}{4g(r)}.  \label{11}
\end{equation}%
It is worth noting that $l$ stands for the rotation parameter of the black
hole, $j$ is known as the Gödel parameter, and $m$ stands for the mass
parameter of the black hole. The zero limit of the $j$ and $l$ reduces the
metric \eqref{1} to the 5-dimensional Schwarzschild black hole. Besides, in the
case of $m=l=0$ the metric reduces to that of 5-dimensional Gödel
universe \cite{gd1}. Metric in Eq. \eqref{1}  is nothing
but the Schwarzschild-Gödel black hole when $l=0$. Furthermore, metric in Eq. \eqref{1}
is regular at the horizons and the tensor scalars become singular only at $%
r=0$.

Employing the Wick-rotation method, it was shown that the Hawking
temperature of the 5DKGBH reads \cite{gd2}

\begin{equation}
T_{H}=\frac{m\left[ (1-8j^{2}m-4ml)r_{+}^{2}-2l^{2}\right] }{\pi r_{+}^{3}%
\sqrt{(1-8j^{2}m)r_{+}^{4}+2ml^{2}-4j^{2}r_{+}^{6}}},  \label{12}
\end{equation}

in which $r_{+}$\ denotes the event horizon of the black hole. In the
following section, we shall attempt to re-derive Eq. \eqref{12}  by considering
quantum tunneling phenomenon of the massive vector particles.

\section{Quantum Tunneling of Spin-1 Particles From 5DKGBH }

In this section, we calculate the Hawking radiation of the spin-1 particles
from the 5DKGBH spacetime. For this purpose, we firstly transform the
lapse-shift form of the 5GKGBH in Eq. \eqref{10} to the Gullstrand-Painlevé coordinates \cite{pan1,pan2}. 
Then, we shall employ the Proca
equation \cite{proca1} with the Hamilton-Jacobi ansatz. Our computations will be in
the framework of semi-classical analysis of the WKB approximation.

Proca equation for the massive spin-1 particles is given by \cite{proca1,kruglov1,kruglov2}

\begin{equation}
\frac{1}{\sqrt{-g}}\partial _{\mu }\left( \sqrt{-g}\,\Psi ^{\nu \mu }\right)
+\frac{m^{2}}{\hbar ^{2}}\Psi ^{\nu }=0,  \label{13}
\end{equation}%
where the anti-symmetric tensor is governed by
\begin{equation}
\Psi _{\mu \nu }=\partial _{\mu }\Psi _{\nu }-\partial _{\nu }\Psi _{\mu },
\label{14}
\end{equation}

by which $\Psi ^{\nu }$ is the vector field. Without loss of generality, one
can simplify the equations to be obtained from Eq. \eqref{13} by redefining the metric in Eq. \eqref{10} with the dragging coordinate \cite{mann4} $\chi =\phi -\Omega _{H}t$ in which $\Omega _{H}$ represents the angular
velocity of locally non-rotating observers:

\begin{equation}
\Omega _{H}=\frac{d\phi }{dt}=\frac{a(r)}{g(r)}.  \label{15}
\end{equation}
We are interested in tunneling of massive vector particles that have no angular momentum ($\ell =0$) so we set $d \chi=0$ (and for convenience also $d\psi =d\theta =0$). Thus, we find out the following simplified metric:

\begin{equation}
ds^{2}=-\frac{r^{2}V(r)}{4g(r)}dt^{2}+\frac{dr^{2}}{V(r)}.  \label{16}
\end{equation}%
To rewrite Eq. \eqref{16} in Gullstrand-Painlevé coordinates, one can use the
following transformation: 
\begin{equation}
t\rightarrow t-\frac{2\sqrt{g(r)}}{rV(r)}\sqrt{1-V(r)}dr.  \label{17}
\end{equation}%
Thus, we have 
\begin{equation}
ds^{2}=-\frac{r^{2}V(r)}{4g(r)}dt^{2}+\frac{r}{\sqrt{g(r)}}\sqrt{1-V(r)}%
drdt+dr^{2},  \label{18}
\end{equation}%
which can be rewritten as follows:

\begin{equation}
ds^{2}=-F(r)dt^{2}+H(r)drdt+dr^{2}  \label{19}
\end{equation}%
where

\begin{equation}
F(r)=\frac{r^{2}V(r)}{4g(r)},  \label{20}
\end{equation}

\begin{equation}
H(r)=\frac{r}{\sqrt{g(r)}}\sqrt{1-V(r)}.  \label{21}
\end{equation}

Let us now assume an ansatz for the vector spinor that suits for the WKB approximation as follows

\begin{equation}
\Psi _{\nu }=C_{\nu }\exp \left( \frac{i}{\hbar }\left( S_{0}(t,r)+\hbar
\,S_{1}(t,r)+\hdots.\right) \right) ,  \label{22}
\end{equation}

where $C_{\nu }=\{C_{1},C_{2}\}$ represents some arbitrary constants and
the leading order action $S_{0}$ is defined as 
\begin{equation}
S_{0}(t,r)=-Et+W(r)+k,  \label{23}
\end{equation}

in which $E$ and $k$ represent the energy and complex constant,
respectively. 
\begin{widetext}
Hence, the resulting equations obtained in the leading order $%
\hbar $ are as follows:
\begin{eqnarray}
\frac{2H(r)m^{2}+4C_{2}F(r)m^{2}-4C_{1}EW'(r)-4C_{2}E^{2}}{H(r)^{2}+4F(r)}
&=&0  \label{24} \\
\frac{-4C_{1}W'(r)^{2}-4C_{2}EW'(r)-4m^{2}\left( -\frac{C_{2}H(r)}{2}%
+C_{1}\right) }{H(r)^{2}+4F(r)} &=&0  \label{25}
\end{eqnarray}
\end{widetext}

Thus, the associated matrix of Eqs. (24) and (25) becomes 

\begin{equation}
\aleph =\left[ 
\begin{array}{cc}
{\frac{2H\left( r\right) {m}^{2}-4EW'\left(
r\right) }{\left( H\left( r\right) \right) ^{2}+4F\left( r\right) }} & {%
\frac{4F\left( r\right) {m}^{2}-4{E}^{2}}{\left( H\left( r\right) \right)
^{2}+4F\left( r\right) }} \\ 
\frac{-4W'(r)^{2}-4m^{2}}{H(r)^{2}+4F(r)} & {\frac{2H\left( r\right) {m}%
^{2}-4EW'\left( r\right) }{\left( H\left(
r\right) \right) ^{2}+4F\left( r\right) }}%
\end{array}%
\right]   \label{26}
\end{equation}

It is fact that when the determinant of matrix of homogeneous system of
linear equations is zero ($\det \aleph =0$) \cite{kruglov1}, we have a non-trivial solution.

\begin{widetext}
Based on this fact, we compute the determinant of the matrix (26) as
\begin{equation}
\frac{%
-4m^{2}(-H(r)^{2}m^{2}+4EH(r)W'(r)-4F(r)W'(r)^{2}-4F(r)m^{2}+4E^{2})}{%
(H(r)^{2}+4F(r))^{2}}=0.  \label{27}
\end{equation}
\end{widetext}

\begin{widetext}
From Eq. (27), one can obtain the following integral solution for the
radial function:
\begin{equation}
W_{\pm }=\pm \int \frac{1}{2F(r)}\left( EH(r)+\sqrt{%
4E^{2}F(r)-F(r)H(r)^{2}m^{2}+E^{2}H(r)^{2}-4F(r)^{2}m^{2}}\right) dr.
\label{28}
\end{equation}
\end{widetext}
Around the event horizon $\left[ F(r_{+})=0\right] $, Eq. (28) reduces to

\begin{equation}
W_{\pm }\approx \pm \int \frac{EH(r)}{F(r)}dr.  \label{29}
\end{equation}

It is clear from Eq. (29) that the integrand possesses a simple pole at the
event horizon. To overcome this difficulty, we use the Feynman's
prescription \cite{feyn}. After a straightforward calculation, one can obtain

\begin{equation}
W_{\pm }\approx \pm i\pi E\frac{H(r_{+})}{F'(r_{+})}.  \label{30n}
\end{equation}

Now, we have two solutions: $W_{+}$ corresponds to the radial solutions of
the outgoing spin-1 particles and $W_{-}$ represents the ingoing particles
that move towards the 5DKGBH. Therefore, the emission and absorption
probabilities of the spin-1 particles crossing the 5DKGBH's event horizon
each way become

\begin{equation}
P_{ems}=\exp \left( -\frac{2}{\hbar }\mathrm{Im}S_{0}\right) =\mathrm{exp}\left[
-\frac{2}{\hbar }(\mathrm{Im}W_{+}+\mathrm{Im}k)\right],  \label{31n}
\end{equation}

\begin{equation}
P_{abs}=\exp \left( -\frac{2}{\hbar }\mathrm{Im}S_{0}\right) =\mathrm{exp}\left[
-\frac{2}{\hbar }(\mathrm{Im}W_{-}+\mathrm{Im}k)\right]. \label{32n}
\end{equation}

According to our understanding of classical black holes, we should set the probability of the ingoing particle to 100\%. This is possible with $%
\mathrm{Im}k=-\mathrm{Im}W_{-}$. On the other hand, we know from Eq. (30)
that $W_{-}=-W_{+}$. Combining those information, one can read the quantum
tunneling rate of the spin-1 particles as

\begin{equation}
\Gamma =\frac{P_{ems}}{P_{abs}}=\mathrm{exp}\left[ -\frac{4}{\hbar }\mathrm{%
Im}W_{+}\right] .  \label{33}
\end{equation}%
Recalling the Boltzmann factor $\Gamma =\exp \left( -\frac{E}{T_{S}}\right) $
\cite{wald}, the surface temperature of the 5DKGBH can be found as

\begin{eqnarray}
T_{S} &=&\frac{F'(r_{+})}{4\pi H(r_{+})},  \notag \\
&=&\frac{m\left[ (1-8j^{2}m-4ml)r_{+}^{2}-2l^{2}\right] }{\pi r_{+}^{3}\sqrt{%
(1-8j^{2}m)r_{+}^{4}+2ml^{2}-4j^{2}r_{+}^{6}}}.  \label{34n}
\end{eqnarray}

One can immediately observe that $T_{S}$ is identical to the Eq. (12), which
is the Hawking temperature of the 5DKGBH computed via the method of Wick rotation.

\section{Conclusion}

To summarize, in this paper, we have studied the Hawking radiation of the
5DKGBH. For this purpose, we have considered the massive spin-1 particles that
are quantum mechanically allowed to emit from the 5DKGBH. We have started our
analysis by considering the Proca equation (13). However, due to the lenghty
Proca equations of the lapse-shift metric of the 5DKGBH (10), we have transformed the metric (10) to the dragging framework expressed in the Gullstrand-Painlevé coordinates (18). Then, with
the help of the Hamilton-Jacobi ansatz of the vector field (22), we have
shown how the quantum tunneling of massive spin-1 particles from the 5DKGBH
results in the Hawking temperature. 
We also plan to extend the current work
to the quantum tunneling of the massive spin-2 particles \cite{a33}
from the 5DKGBH. This will be our next study in the near future.

\begin{acknowledgments}
The authors are grateful to the Editor and anonymous Referees for their
valuable comments and suggestions to improve the paper. This work was supported by the Chilean FONDECYT Grant No. 3170035 (A\"{O}).
\end{acknowledgments}

\end{document}